\documentclass[11pt,a4paper]{article}
\usepackage[hyperref]{emnlp-ijcnlp-2019}
\usepackage{times}
\usepackage{latexsym}
\usepackage{multirow}
\usepackage{url}
\usepackage{enumitem}   

\usepackage{graphicx}
\usepackage{amsmath}
\usepackage{stmaryrd}
\usepackage{amsmath,amssymb,amsfonts}
\usepackage{multicol,lipsum}
\usepackage{graphicx}
\usepackage{subcaption}

\usepackage{url}

\aclfinalcopy 

\title{A Hierarchical Location Prediction Neural Network for Twitter User Geolocation}

\author{Binxuan Huang \\
  School of Computer Science \\
  Carnegie Mellon University \\
  \texttt{binxuanh@cs.cmu.edu} \\\And
  Kathleen M. Carley \\
  School of Computer Science \\
  Carnegie Mellon University \\
  \texttt{kathleen.carley@cs.cmu.edu} \\}

\date{}

\begin{document}
\maketitle
\begin{abstract}
    Accurate estimation of user location is important for many online services. Previous neural network based methods largely ignore the hierarchical structure among locations. In this paper, we propose a hierarchical location prediction neural network for Twitter user geolocation. Our model first predicts the home country for a user, then uses the country result to guide the city-level prediction. In addition, we employ a character-aware word embedding layer to overcome the noisy information in tweets. With the feature fusion layer, our model can accommodate various feature combinations and achieves state-of-the-art results over three commonly used benchmarks under different feature settings. It not only improves the prediction accuracy but also greatly reduces the mean error distance.
\end{abstract}
\section{Introduction}
Accurate estimation of user location is an important factor for many online services, such as recommendation systems \cite{quercia2010recommending}, event detection \cite{sakaki2010earthquake}, and disaster management \cite{carley2016crowd}. Though internet service providers can directly obtain users' location information from some explicit metadata like IP address and GPS signal, such private information is not available for third-party contributors. With this motivation, researchers have developed location prediction systems for various platforms, such as Wikipedia \cite{overell2009geographic}, Facebook \cite{backstrom2010find}, and Twitter \cite{han2012geolocation}.

In the case of Twitter, due to the sparsity of geotagged tweets \cite{graham2014world} and the unreliability of user self-declared home location in profile \cite{hecht2011tweets}, there is a growing body of research trying to determine users' locations automatically. Various methods have been proposed for this purpose. They can be roughly divided into three categories. The first type consists of tweet text-based methods, where the word distribution is used to estimate geolocations of users \cite{roller2012supervised, wing2011simple}. In the second type, methods combining metadata features such as time zone and profile description are developed to improve performance \cite{han2013stacking}. Network-based methods form the last type. Several studies have shown that incorporating friends' information is very useful for this task \cite{miura2017unifying, ebrahimi2018unified}. Empirically, models enhanced with network information work better than the other two types, but they do not scale well to larger datasets \cite{rahimi2015twitter}.

In recent years, neural network based prediction methods have shown great success on this Twitter user geolocation prediction task \cite{ rahimi2017neural, miura2017unifying}. However, these neural network based methods largely ignore the hierarchical structure among locations (eg. country versus city), which have been shown to be very useful in previous study \cite{mahmud2012tweet, wing2014hierarchical}. In recent work, \citet{huang2017predicting} also demonstrate that country-level location prediction is much easier than city-level location prediction. It is natural to ask whether we can incorporate the hierarchical structure among locations into a neural network and use the coarse-grained location prediction to guide the fine-grained prediction. Besides, most of these previous work uses word-level embeddings to represent text, which may not be sufficient for noisy text from social media.

In this paper, we present a hierarchical location prediction neural network (HLPNN) for user geolocation on Twitter. Our model combines text features, metadata features (personal description, profile location, name, user language, time zone), and network features together.  It uses a character-aware word embedding layer to deal with the noisy text and capture out-of-vocabulary words. With transformer encoders, our model learns the correlation between different feature fields and outputs two classification representations for country-level and city-level predictions respectively. It first computes the country-level prediction, which is further used to guide the city-level prediction. Our model is flexible in accommodating different feature combinations, and it achieves state-of-the-art results under various feature settings. 
\section{Related Work}
Because of insufficient geotagged data \cite{graham2014world, binxuan2019large}, there is a growing interest in predicting Twitter users' locations. Though there are some potential privacy concerns, user geolocation is a key factor for many important applications such as earthquake detection \cite{earle2012twitter}, and disaster management \cite{carley2016crowd}, health management \cite{huang2018location}.

Early work tried to identify users' locations by mapping their IP addresses to physical locations \cite{buyukokkten1999exploiting}. However, such private information is only accessible to internet service providers. There is no easy way for a third-party to find Twitter users' IP addresses. Later, various text-based location prediction systems were proposed. \citet{bilhaut2003geographic} utilize a geographical gazetteer as an external lexicon and present a rule-based geographical references recognizer. \citet{amitay2004web} extracted location-related information listed in a gazetteer from web content to identify geographical regions of webpages. However, as shown in \cite{berggren2016inferring}, performances of gazetteer-based methods are hindered by the noisy and informal nature of tweets.

Moving beyond methods replying on external knowledge sources (e.g. IP and gazetteers), many machine learning based methods have recently been applied to location prediction. Typically, researchers first represent locations as earth grids \cite{wing2011simple, roller2012supervised}, regions \cite{miyazaki2018twitter, qian2017probabilistic}, or cities \cite{han2013stacking}. Then location classifiers are built to categorize users into different locations. \citet{han2012geolocation} first utilized feature selection methods to find location indicative words, then they used multinomial naive Bayes and logistic regression classifiers to find correct locations. \citet{han2013stacking} further present a stacking based method that combines tweet text and metadata together. Along with these classification methods, some approaches also try to learn topic regions automatically by topic modeling, but these do not scale well to the magnitude of social media \cite{hong2012discovering, zhang2017rate}.

Recently, deep neural network based methods are becoming popular for location prediction \cite{miura2016simple}. \citet{huang2017predicting} integrate text and user profile metadata into a single model using convolutional neural networks, and their experiments show superior performance over stacked naive Bayes classifiers. \citet{miura2017unifying, ebrahimi2018unified} incorporate user network connection information into their neural models, where they use network embeddings to represent users in a social network. \citet{rahimi2018semi} also uses text and network feature together, but their approach is based on graph convolutional neural networks.

Similar to our method, some research has tried to predict user location hierarchically \cite{mahmud2012tweet, wing2014hierarchical}. \citet{mahmud2012tweet} develop a two-level hierarchical location classifier which first predicts a coarse-grained location (country, time zone), and then predicts the city label within the corresponding coarse region. \citet{wing2014hierarchical} build a hierarchical tree of earth grids. The probability of a final fine-grained location can be computed recursively from the root node to the leaf node. Both methods have to train one classifier separately for each parent node, which is quite time-consuming for training deep neural network based methods. Additionally, certain coarse-grained locations may not have enough data samples to train a local neural classifier alone. Our hierarchical location prediction neural network overcomes these issues and only needs to be trained once.
\section{Method}
There are seven features we want to utilize in our model --- tweet text, personal description, profile location, name, user language, time zone, and mention network. The first four features are text fields where users can write anything they want. User language and time zone are two categorical features that are selected by users in their profiles. Following previous work \cite{rahimi2018semi}, we construct mention network directly from mentions in tweets, which is also less expensive to collect than following network\footnote{https://developer.twitter.com}.


\begin{figure*}[!h]
    \centering
    \includegraphics[width=0.8\textwidth]{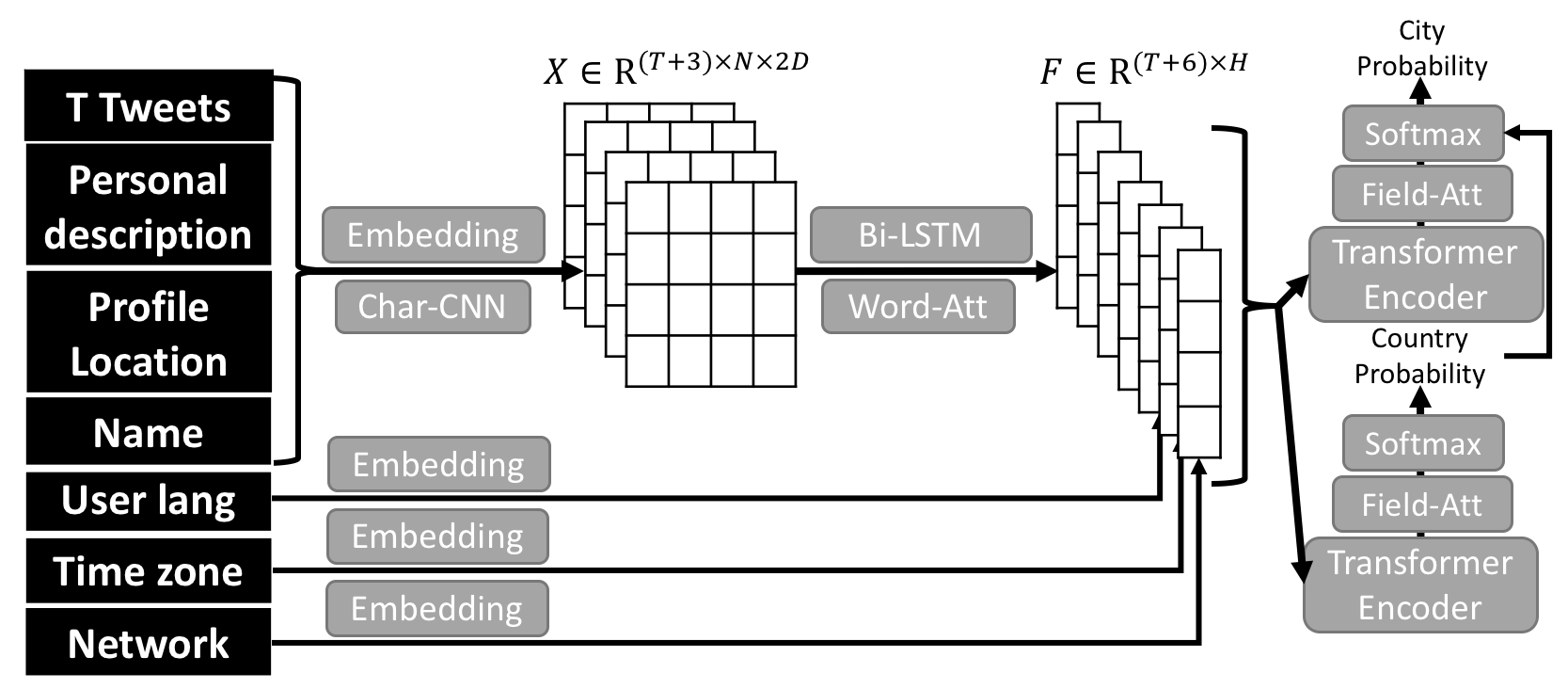}
    \caption{The architecture of our hierarchical location prediction neural network.}
    \label{arch}
\end{figure*}

The overall architecture of our hierarchical location prediction model is shown in Figure \ref{arch}. It first maps four text features into a word embedding space. A bidirectional LSTM (Bi-LSTM) neural network \cite{hochreiter1997long} is used to extract location-specific features from these text embedding vectors. Following Bi-LSTM, we use a word-level attention layer to generate representation vectors for these text fields. Combining all the text representations, a user language embedding, a timezone embedding, and a network embedding, we apply several layers of transformer encoders \cite{vaswani2017attention} to learn the correlation among all the feature fields. The probability for each country is computed after a field-level attention layer. Finally, we use the country probability as a constraint for the city-level location prediction. We elaborate details of our model in the following sections.

\subsection{Word Embedding}
Assume one user has $T$ tweets, there are $T+3$ text fields for this user including personal description, profile location, and name. We first map each word in these $T+3$ text fields into a low dimensional embedding space. The embedding vector for word $w$ is computed as $x_w = [E(w),CNN_c(w)]$, where $[,]$ denotes vector concatenation. $E(w)$ is the word-level embedding retrieved directly from an Embedding matrix $E\in R^{V\times D}$ by a lookup operation, where $V$ is the vocabulary size, and $D$ is the word-level embedding dimension. $CNN_c(w)$ is a character-level word embedding that is generated from a character-level convolutional layer. Using character-level word embeddings is helpful for dealing with out-of-vocabulary tokens and overcoming the noisy nature of tweet text.

The character-level word embedding generation process is as follows. For a character $c_i$ in the word $w=(c_1,...,c_k)$, we map it into a character embedding space and get a vector $v_{c_i}\in R^{d}$. In the convolutional layer, each filter $u \in R^{l_c \times d}$ generates a feature vector $\boldsymbol{ \theta }=[\theta_1,\theta_2,...,\theta_{k-l_c+1}]\in R^{k-l_c+1}$, where $\theta_i=\textit{relu}(u \circ v_{c_i:c_{i+l_c-1}}+b)$. $b$ is a bias term, and ``$\circ$'' denotes element-wise inner product between $u$ and character window $v_{c_i:c_{i+l_c-1}}\in R^{l_c\times d}$. After this convolutional operation, we use a max-pooling operation to select the most representative feature $\hat \theta = max(\boldsymbol{ \theta })$. With $D$ such filters, we get the character-level word embedding $CNN_c(w)\in R^D$.

\subsection{Text Representation}

 After the word embedding layer, every word in these $T+3$ texts are transformed into a $2D$ dimension vector. Given a text with word sequence $(w_1,...,w_N)$, we get a word embedding matrix $X\in R^{N\times 2D}$ from the embedding layer. We then apply a Bi-LSTM neural network to extract high-level semantic representations from text embedding matrices. 

At every time step $i$, a forward LSTM takes the word embedding $x_i$ of word $w_i$ and previous state $\overrightarrow {h_{i-1}}$ as inputs, and generates the current hidden state $\overrightarrow h_i$. A backward LSTM reads the text from $w_N$ to $w_1$ and generates another state sequence. The hidden state $h_i \in R^{2D}$ for word $w_i$ is the concatenation of $\overrightarrow {h_i}$ and $\overleftarrow {h_i}$. Concatenating all the hidden states, we get a semantic matrix $H\in R^{N\times 2D}$
\begin{equation*}
    \begin{aligned}
   & \overrightarrow {h_i}=\overrightarrow{LSTM}(x_i,\overrightarrow {h_{i-1}})\\
    &\overleftarrow {h_i}=\overleftarrow{LSTM}(x_i,\overleftarrow {h_{i+1}}) \\
    \end{aligned}
\end{equation*}

Because not all words in a text contribute equally towards location prediction, we further use a multi-head attention layer \cite{vaswani2017attention} to generate a representation vector $f\in R^{2D}$ for each text. There are $h$ attention heads that allow the model to attend to important information from different representation subspaces. Each head computes a text representation as a weighted average of these word hidden states. The computation steps in a multi-head attention layer are as follows.
\begin{equation*}
    \begin{aligned}
    & f = \operatorname{MultiHead}(q,H)=[\operatorname{head_1},...,\operatorname{head_h}]W^O \\
    & \operatorname{head_i}(q,H) = \operatorname{softmax}(\frac{qW_i^Q\cdot (HW_i^K)^T}{\sqrt{d_k}})HW_i^V
    \end{aligned}
\end{equation*}
where $q\in R^{2d}$ is an attention context vector learned during training, $W_i^Q,W_i^K,W_i^V\in R^{2D\times d_k}$, and $W^O \in R^{2D\times 2D}$ are projection parameters, $d_k=2D/h$. An attention head $head_i$ first projects the attention context $q$ and the semantic matrix $H$ into query and key subspaces by $W_i^Q$, $W_i^K$ respectively. The matrix product between query $qW_i^Q$ and key $HW_i^K$ after softmax normalization is an attention weight that indicates important words among the projected value vectors $HW_i^V$. Concatenating $h$ heads together, we get one representation vector $f\in R^{2D}$ after projection by $W^O$ for each text field.

\subsection{Feature Fusion}
For two categorical features, we assign an embedding vector with dimension $2D$ for each time zone and language. These embedding vectors are learned during training. We pretrain network embeddings for users involved in the mention network using LINE \cite{tang2015line}. Network embeddings are fixed during training. We get a feature matrix $F\in R^{(T+6)\times 2D}$ by concatenating text representations of $T+3$ text fields, two embedding vectors of categorical features, and one network embedding vector.

We further use several layers of transformer encoders \cite{vaswani2017attention} to learn the correlation between different feature fields. Each layer consists of a multi-head self-attention network and a feed-forward network (FFN). One transformer encoder layer first uses input feature to attend important information in the feature itself by a multi-head attention sub-layer. Then a linear transformation sub-layer $FFN$ is applied to each position identically. Similar to \citet{vaswani2017attention}, we employ residual connection \cite{he2016deep} and layer normalization \cite{ba2016layer} around each of the two sub-layers. The output $F_1$ of the first transformer encoder layer is generated as follows.
\begin{equation*}
    \begin{aligned}
        & F' = \operatorname{LayerNorm}(\operatorname{MultiHead}(F,F)+F) \\
        &F_1 = \operatorname{LayerNorm}(\operatorname{FFN}(F')+F')\\
    \end{aligned}
\end{equation*}
where $\operatorname{FFN}(F')=\operatorname{max}(0,F'W_1+b_1)W_2+b2$, $W_1\in R^{2D\times D_{ff}}$, and $W_2\in R^{D_{ff}\times 2D}$. 

Since there is no position information in the transformer encoder layer, our model cannot distinguish between different types of features, eg. tweet text and personal description. To overcome this issue, we add feature type embeddings to the input representations $F$. There are seven features in total. Each of them has a learned feature type embedding with dimension $2D$ so that one feature type embedding and the representation of the corresponding feature can be summed. 

Because the input and the output of transformer encoder have the same dimension, we stack $L$ layers of transformer encoders to learn representations for country-level prediction and city-level prediction respectively. These two sets of encoders share the same input $F$, but generate different representations $F_{co}^L$ and $F_{ci}^L$ for country and city predictions. 

The final classification features for country-level and city-level location predictions are the row-wise weighted average of $F_{co}$ and $F_{ci}$. Similar to the word-level attention, we use a field-level multi-head attention layer to select important features from $T+6$ vectors and fuse them into a single vector.  
\begin{equation*}
    \begin{aligned}
    F_{co} = \operatorname{MultiHead}(q_{co},F_{co}^L)\\
    F_{ci} = \operatorname{MultiHead}(q_{ci},F_{ci}^L)
    \end{aligned}
\end{equation*}
 where $q_{co}, q_{ci}\in R^{2D}$ are two attention context vectors.

\begin{table*}[!h]
\centering
\resizebox{0.8\textwidth}{!}{
\begin{tabular}{lccccccccc}
\hline
 \hline
\multirow{2}{*}{}                                             & \multicolumn{3}{c}{Twitter-US} & \multicolumn{3}{c}{Twitter-World} & \multicolumn{3}{c}{WNUT} \\ \cline{2-10} 
                                                              & Train    & Dev.     & Test     & Train     & Dev.      & Test      & Train  & Dev.   & Test   \\ \hline
\# users                                                      & 429K     & 10K      & 10K      & 1.37M     & 10K       & 10K       & 742K   & 7.46K  & 10K    \\ \hline
\begin{tabular}[c]{@{}l@{}}\# users \\ with meta\end{tabular} & 228K     & 5.32K    & 5.34K    & 917K      & 6.50K     & 6.48K     & 742K   & 7.46K  & 10K    \\ \hline
\# tweets                                                     & 36.4M    & 861K     & 831K     & 11.2M     & 488K      & 315K      & 8.97M  & 90.3K  & 99.7K  \\ \hline
\begin{tabular}[c]{@{}l@{}} \# tweets \\ per user\end{tabular} & 84.60    & 86.14    & 83.12    & 8.16      & 48.83     & 31.59     & 12.09  & 12.10  & 9.97   \\ \hline
 \hline
\end{tabular}
}
\caption{A brief summary of our datasets. For each dataset, we report the number of users, number of users with metadata, number of tweets, and average number of tweets per user. We collected metadata for 53\% and 67\% of users in Twitter-US and Twitter-World. Time zone information was not available when we collected metadata for these two datasets. About 25\% of training and development users' data was inaccessible when we collected WNUT in 2017.}
\label{data}
\end{table*}

\subsection{Hierarchical Location Prediction}
 The final probability for each country is computed by a softmax function
\begin{equation*}
    \begin{aligned}
        P_{co} = \operatorname{softmax}(W_{co}F_{co}+b_{co})
    \end{aligned}
\end{equation*}
where $W_{co} \in R^{M_{co}\times 2D}$ is a linear projection parameter, $b_{co}\in R^{M_{co}}$ is a bias term, and $M_{co}$ is the number of countries. 

After we get the probability for each country, we further use it to constrain the city-level prediction 
\begin{equation*}
    \begin{aligned}
        P_{ci} = &\operatorname{softmax}(W_{ci}F_{ci}+b_{ci} + \lambda P_{co} Bias) 
    \end{aligned}
\end{equation*}
where $W_{ci} \in R^{M_{ci}\times 2D}$ is a linear projection parameter, $b_{ci}\in R^{M_{ci}}$ is a bias term, and $M_{ci}$ is the number of cities. $Bias\in R^{M_{co}\times M_{ci}}$ is the country-city correlation matrix. If city $j$ belongs to country $i$, then $Bias_{ij}$ is $0$, otherwise $-1$. $\lambda$ is a penalty term learned during training. The larger of $\lambda$, the stronger of the country constraint. In practise, we also experimented with letting the model learn the country-city correlation matrix during training, which yields similar performance.

We minimize the sum of two cross-entropy losses for country-level prediction and city-level prediction. 
\begin{equation*}
    \begin{aligned}
        loss = -( Y_{ci}\cdot log P_{ci} + \alpha Y_{co}\cdot log P_{co})
    \end{aligned}
\end{equation*}
where $Y_{ci}$ and $Y_{co}$ are one-hot encodings of city and country labels. $\alpha$ is the weight to control the importance of country-level supervision signal. Since a large $\alpha$ would potentially interfere with the training process of city-level prediction, we just set it as $1$ in our experiments. Tuning this parameter on each dataset may further improve the performance. 

\section{Experiment Settings}
\subsection{Datasets}

To validate our method, we use three widely adopted Twitter location prediction datasets. Table \ref{data} shows a brief summary of these three datasets. They are listed as follows.

 \textbf{Twitter-US} is a dataset compiled by \citet{roller2012supervised}. It contains 429K training users, 10K development users, and 10K test users in North America. The ground truth location of each user is set to the first geotag of this user in the dataset. We assign the closest city to each user's ground truth location using the city category built by \citet{han2012geolocation}. Since this dataset only covers North America, we change the first level location prediction from countries to administrative regions (eg. state or province). The administrative region for each city is obtained from the original city category.

\textbf{Twitter-World} is a Twitter dataset covering the whole world, with 1,367K training users, 10K development users, and 10K test users \cite{han2012geolocation}. The ground truth location for each user is the center of the closest city to the first geotag of this user. Only English tweets are included in this dataset, which makes it more challenging for a global-level location prediction task.

We downloaded these two datasets from Github \footnote{https://github.com/afshinrahimi/geomdn}. Each user in these two datasets is represented by the concatenation of their tweets, followed by the geo-coordinates. We queried Twitter's API to add user metadata information to these two datasets in February 2019. We only get metadata for about 53\% and 67\% users in Twitter-US and Twitter-World respectively. Because of Twitter's privacy policy change, we could not get the time zone information anymore at the time of collection. 

 \textbf{WNUT} was released in the 2nd Workshop on Noisy User-generated Text \cite{han2016twitter}. The original user-level dataset consists of 1 million training users, 10K users in development set and test set each. Each user is assigned with the closest city center as the ground truth label. Because of Twitter's data sharing policy, only tweet IDs of training and development data are provided. We have to query Twitter's API to reconstruct the training and development dataset. We finished our data collection around August 2017. About 25\% training and development users' data cannot be accessed at that time. The full anonymized test data is downloaded from the workshop website \footnote{https://noisy-text.github.io/2016/geo-shared-task.html}.

\subsection{Text Preprocessing \& Network Construction}

For all the text fields, we first convert them into lower case, then use a tweet-specific tokenizer from NLTK\footnote{https://www.nltk.org/api/nltk.tokenize.html} to tokenize them. To keep a reasonable vocabulary size, we only keep tokens with frequencies greater than 10 times in our word vocabulary. Our character vocabulary includes characters that appear more than 5 times in the training corpus.

We construct user networks from mentions in tweets. For WNUT, we keep users satisfying one of the following conditions in the mention network: (1) users in the original dataset (2) users who are mentioned by two different users in the dataset. For Twitter-US and Twitter-World, following previous work \cite{rahimi2018semi}, a uni-directional edge is set if two users in our dataset directly mentioned each other, or they co-mentioned another user. We remove celebrities who are mentioned by more than 10 different users from the mentioning network. These celebrities are still kept in the dataset and their network embeddings are set as 0.

\subsection{Evaluation Metrics}
We evaluate our method using four commonly used metrics listed below.\\
\textbf{Accuracy}: The percentage of correctly predicted home cities.\\
\textbf{Acc@161}: The percentage of predicted cities which are within a 161 km (100 miles) radius of true locations to capture near-misses.\\
\textbf{Median}: The median distance measured in kilometer from the predicted city to the true location coordinates. \\
\textbf{Mean}: The mean value of error distances in predictions.

\subsection{Hyperparameter Settings}
In our experiments, we initialize word embeddings with released 300-dimensional Glove vectors \cite{pennington2014glove}. For words not appearing in Glove vocabulary, we randomly initialize them from a uniform distribution U(-0.25, 0.25). We choose the character embedding dimension as 50. The character embeddings are randomly initialized from a uniform distribution U(-1.0,1.0), as well as the timezone embeddings and language embeddings. These embeddings are all learned during training. Because our three datasets are sufficiently large to train our model, the learning is quite stable and performance does not fluctuate a lot.

Network embeddings are trained using LINE \cite{tang2015line} with parameters of dimension 600, initial learning rate 0.025, order 2, negative sample size 5, and training sample size 10000M. Network embeddings are fixed during training. For users not appearing in the mention network, we set their network embedding vectors as $0$.

\begin{table}[!h]
\resizebox{0.48\textwidth}{!}{
\begin{tabular}{lccc}
\hline
\hline
                                                                              & Twitter-US & Twitter-World & WNUT      \\ \hline
Batch size                                                                    & 32         & 64            & 64        \\ \hline
Initial learning rate                                                         & $10^{-4}$  & $10^{-4}$     & $10^{-4}$ \\ \hline
\begin{tabular}[c]{@{}l@{}}$D$: Word embedding \\ dimension\end{tabular}      & 300        & 300           & 300       \\ \hline
\begin{tabular}[c]{@{}l@{}}$d$: Char. embedding\\  dimension\end{tabular}     & 50         & 50            & 50        \\ \hline
\begin{tabular}[c]{@{}l@{}}$l_c$: filter sizes\\ in Char. CNN\end{tabular}    & 3,4,5      & 3,4,5         & 3,4,5     \\ \hline
\begin{tabular}[c]{@{}l@{}}Filter number \\ for each size\end{tabular}        & 100        & 100           & 100       \\ \hline
$h$: number of heads                                                          & 10         & 10            & 10        \\ \hline
\begin{tabular}[c]{@{}l@{}}$L$: layers of \\ transformer encoder\end{tabular} & 3          & 3             & 3         \\ \hline
$\lambda$: initial penalty term                                              & 1         & 1            & 1        \\ \hline
\begin{tabular}[c]{@{}l@{}}$\alpha$: weight for country\\ supervision \end{tabular}  & 1         & 1            & 1        \\ \hline
\begin{tabular}[c]{@{}l@{}}$D_{ff}$: inner \\ dimension of FFN\end{tabular}   & 2400       & 2400          & 2400      \\ \hline
\begin{tabular}[c]{@{}l@{}}Max number of \\ tweets per user\end{tabular}      & 100        & 50           & 20        \\ \hline \hline

\end{tabular}
}
\caption{A summary of hyperparameter settings of our model.}
\label{parameters}
\end{table}

\begin{table*}[!t]
\resizebox{\textwidth}{!}{
\begin{tabular}{lcccccccccc}
\hline
\hline
\multirow{2}{*}{}             & \multicolumn{3}{c}{Twitter-US} & \multicolumn{3}{c}{Twitter-World} & \multicolumn{4}{c}{WNUT}                                         \\ \cline{2-11} 
                              &Acc@161$\uparrow$& Median$\downarrow$ & Mean$\downarrow$ & Acc@161$\uparrow$ & Median$\downarrow$ & Mean$\downarrow$ & Accuracy$\uparrow$ & Acc@161$\uparrow$ & Median$\downarrow$ & Mean$\downarrow$ \\ \hline
Text                          & \multicolumn{10}{c}{}                                                                                                                 \\
\citet{wing2014hierarchical}   & 49.2      & 170.5    & 703.6   & 32.7       & 490.0     & 1714.6   & -             & -             & -              & -               \\
\citet{rahimi2015exploiting}*  & 50        & 159      & 686     & 32         & 530       & 1724     & -             & -             & -              & -               \\
\citet{miura2017unifying}-TEXT & 55.6      & 110.5    & 585.1   & -          & -         & -        & 35.4          & 50.3          & 155.8          & 1592.6          \\
\citet{rahimi2017neural}       & 55        & 91       & 581     & 36         & 373       & 1417     & -             & -             & -              & -               \\
HLPNN-Text                    &\textbf{57.1}&\textbf{89.92}& \textbf{516.6}   &\textbf{40.1}&\textbf{299.1} & \textbf{1048.1}& \textbf{37.3} & \textbf{52.9} & \textbf{109.3} & \textbf{1289.4} \\ \hline
Text+Meta                     &           &          &         &            &           &          &               &               &                &                 \\
\citet{miura2017unifying}-META &\textbf{67.2} & \textbf{46.8} &\textbf{356.3}& -          & -         & -        & 54.7          & 70.2          & 0              & 825.8           \\
HLPNN-Meta                    &  61.1     & 64.3     & 454.8         & \textbf{56.4}&\textbf{86.2}&\textbf{762.1}& \textbf{57.2} & \textbf{73.1} & \textbf{0}     & \textbf{572.5}  \\ \hline
Text+Net                      & \multicolumn{10}{c}{}                                                                                                                 \\
\citet{rahimi2015twitter}*     & 60        & 78       & 529     & 53         & 111       & 1403     & -             & -             & -              & -               \\
\citet{rahimi2017neural}       & 61        & 77       & 515     & 53         & 104       & 1280     & -             & -             & -              & -               \\
\citet{miura2017unifying}-UNET & 61.5      & 65       & 481.5   & -          & -         & -        & \textbf{38.1} & \textbf{53.3} & \textbf{99.9}  & 1498.6          \\
\citet{do2017multiview}       & 66.2       & 45       & 433    & 53.3        & 118       & 1044     & -             &  -            &  -             &   -      \\
\citet{rahimi2018semi}-MLP-TXT+NET & 66   & 56       & 420    & 58          & \textbf{53} & 1030     & -             &  -            &  -             &   -      \\
\citet{rahimi2018semi}-GCN     & 62       & 71       & 485    & 54          & 108        & 1130     & -             &  -            &  -             &   -      \\
HLPNN-Net                     &\textbf{70.8}&\textbf{31.6}&\textbf{361.5} &\textbf{58.9} & 59.9 & \textbf{827.6} & 37.8          & \textbf{53.3} & 105.26         & \textbf{1297.7} \\ \hline
Text+Meta+Net                 &           &          &         &            &           &          &               &               &                &                 \\
\citet{miura2016simple}        & -         & -        & -       & -          & -         & -        & 47.6          & -             & 16.1           & 1122.3          \\
\citet{jayasinghe2016csiro}    & -         & -        & -       & -          & -         & -        & 52.6          & -             & 21.7           & 1928.8          \\
\citet{miura2017unifying}      & 70.1      & 41.9     & 335.7   & -          & -         & -        & 56.4          & 71.9          & \textbf{0}    & 780.5           \\
HLPNN                         & \textbf{72.7} &\textbf{28.2}& \textbf{323.1} &\textbf{68.4} & \textbf{6.20}  & \textbf{610.0} &  \textbf{57.6} & \textbf{73.4} & \textbf{0}  & \textbf{538.8}   \\ \hline 
\hline
\end{tabular}
}
\caption{Comparisons between our method and baselines. We report results under four different feature settings: Text, Text+Metadata, Text+Network, Text+Metadata+Network. ``-'' signifies that no results were published for the given dataset, ``*'' denotes that results are cited from \citet{rahimi2017neural}. Note that \citet{miura2017unifying} only used 279K users added with metadata in their experiments of Twitter-US.}
\label{result}
\end{table*}

A brief summary of hyperparameter settings of our model is shown in Table \ref{parameters}. The initial learning rate is $10^{-4}$. If the validation accuracy on the development set does not increase, we decrease the learning rate to $10^{-5}$ and train the model for additional 3 epochs. Empirically, training terminates within 10 epochs. Penalty $\lambda$ is initialized as $1.0$ and is adapted during training. We apply dropout on the input of Bi-LSTM layer and the output of two sub-layers in transformer encoders with dropout rate 0.3 and 0.1 respectively. We use the Adam update rule \cite{kingma2014adam} to optimize our model. Gradients are clipped between -1 and 1. The maximum numbers of tweets per user for training and evaluating on Twitter-US are 100 and 200 respectively. We only tuned our model, learning rate, and dropout rate on the development set of WNUT.

\section{Results}
\subsection{Baseline Comparisons}

In our experiments, we evaluate our model under four different feature settings: Text, Text+Meta, Text+Network, Text+Meta+Network. HLPNN-Text is our model only using tweet text as input. HLPNN-Meta is the model that combines text and metadata (description, location, name, user language, time zone). HLPNN-Net is the model that combines text and mention network. HLPNN is our full model that uses text, metadata, and mention network for Twitter user geolocation. 

We present comparisons between our model and previous work in Table \ref{result}. As shown in the table, our model outperforms these baselines across three datasets under various feature settings.

Only using text feature from tweets, our model HLPNN-Text works the best among all these text-based location prediction systems and wins by a large margin. It not only improves prediction accuracy but also greatly reduces mean error distance. Compared with a strong neural model equipped with local dialects \cite{rahimi2017neural}, it increases Acc@161 by an absolute value 4\% and reduces mean error distance by about 400 kilometers on the challenging Twitter-World dataset, without using any external knowledge. Its mean error distance on Twitter-World is even comparable to some methods using network feature \cite{do2017multiview}.

With text and metadata, HLPNN-Meta correctly predicts locations of 57.2\% users in WNUT dataset, which is even better than these location prediction systems that use text, metadata, and network. Because in the WNUT dataset the ground truth location is the closest city's center, Our model achieves 0 median error when its accuracy is greater than 50\%. Note that \citet{miura2017unifying} used 279K users added with metadata in their experiments on Twitter-US, while we use all 449K users for training and evaluation, and only 53\% of them have metadata, which makes it difficult to make a fair comparison. 

Adding network feature further improves our model's performances. It achieves state-of-the-art results combining all features on these three datasets. Even though unifying network information is not the focus of this paper, our model still outperforms or has comparable results to some well-designed network-based location prediction systems like \cite{rahimi2018semi}. On Twitter-US dataset, our model variant HLPNN-Net achieves a 4.6\% increase in Acc@161 against previous state-of-the-art methods \cite{do2017multiview} and \cite{rahimi2018semi}. The prediction accuracy of HLPNN-Net on WNUT dataset is similar to \cite{miura2017unifying}, but with a noticeable lower mean error distance. 

\subsection{Ablation Study}
In this section, we provide an ablation study to examine the contribution of each model component. Specifically, we remove the character-level word embedding, the word-level attention, the field-level attention, the transformer encoders, and the country supervision signal one by one at a time. We run experiments on the WNUT dataset with text features.

\begin{table}[!h]
\centering
\resizebox{0.48\textwidth}{!}{
\begin{tabular}{lcccc}
\hline\hline
                         & Accuracy & Acc@161  & Median & Mean   \\ \hline
HLPNN                    & 37.3     & 52.9      & 109.3 & 1289.4  \\ 
w/o Char-CNN             & 36.3     & 51.0     & 130.8 & 1429.9  \\ 
w/o Word-Att             & 36.4    & 51.5    & 130.2 & 1377.5  \\ 
w/o Field-Att            & 37.0     & 52.0      & 121.8 & 1337.5  \\ 
w/o encoders             & 36.8     & 52.5      & 117.4 & 1402.9  \\ 
w/o country              & 36.7     & 52.6      & 124.8 & 1399.2 \\ \hline\hline
\end{tabular}
}
\caption{An ablation study on WNUT dataset.}
\label{ablation}
\end{table}

The performance breakdown for each model component is shown in Table \ref{ablation}. Compared to the full model, we can find that the character-level word embedding layer is especially helpful for dealing with noisy social media text. The word-level attention also provides performance gain, while the field-level attention only provides a marginal improvement. The reason could be the multi-head attention layers in the transformer encoders already captures important information among different feature fields. These two transformer encoders learn the correlation between features and decouple these two level predictions. Finally, using the country supervision can help model to achieve a better performance with a lower mean error distance.

\subsection{Country Effect}
To directly measure the effect of adding country-level supervision, we define a relative country error which is the percentage of city-level predictions located in incorrect countries among all misclassified city-level predictions.
\begin{align*}
\resizebox{0.48\textwidth}{!}{$\operatorname{relative\ country\ error} = \frac{\operatorname{\#\ of\ incorrect\ country}}{\operatorname{\#\ of\ incorrect\ city}}$}
\end{align*}
The lower this metric means the better one model can predict the city-level location, at least in the correct country.

\begin{figure}[!h]
    \centering
    \includegraphics[width=0.5\textwidth]{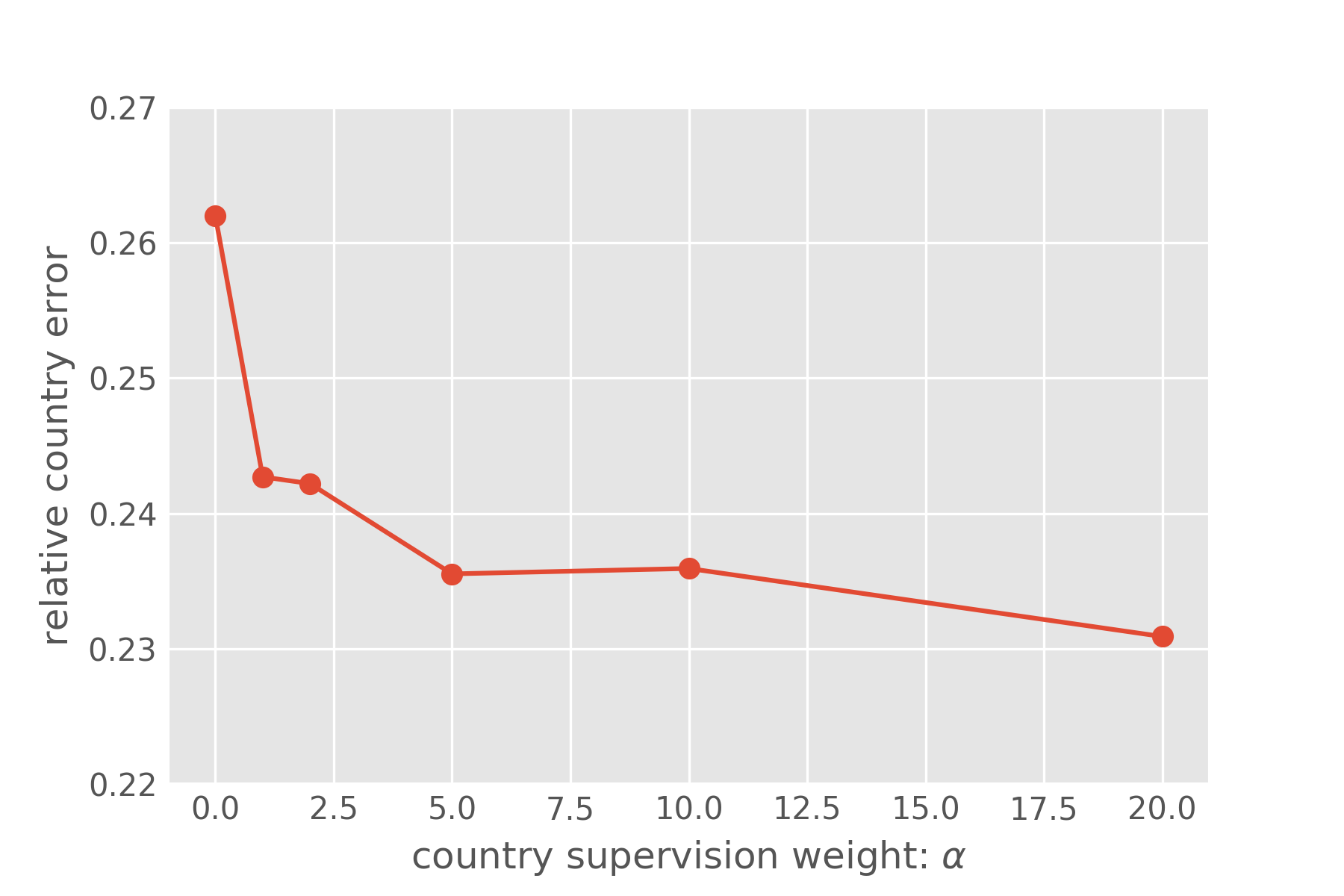}
    \caption{Relative country error with varying $\alpha$ on test dataset. Experiments were conducted on WNUT dataset with text feature.}
    \label{alpha}
\end{figure}

We vary the weight $\alpha$ of country-level supervision signal in our loss function from 0 to 20. The larger $\alpha$ means the more important the country-level supervision during the optimization. When $\alpha$ equals 0, there is no country-level supervision in our model. As shown in Figure \ref{alpha}, increasing $\alpha$ would improve the relative country error from 26.2\% to 23.1\%, which shows the country-level supervision signal indeed can help our model predict the city-level location towards the correct country. This possibly explains why our model has a lower mean error distance when compared to other methods.

\section{Conclusion}

In this paper, we propose a hierarchical location prediction neural network, which combines text, metadata, network information for user location prediction. Our model can accommodate various feature combinations. Extensive experiments have been conducted to validate the effectiveness of our model under four different feature settings across three commonly used benchmarks. Our experiments show our HLPNN model achieves state-of-the-art results on these three datasets. It not only improves the prediction accuracy but also significantly reduces the mean error distance. In our ablation analysis, we show that using character-aware word embeddings is helpful for overcoming noise in social media text. The transformer encoders effectively learn the correlation between different features and decouple the two different level predictions. In our experiments, we also analyzed the effect of adding country-level regularization. The country-level supervision could effectively guide the city-level prediction towards the correct country, and reduce the errors where users are misplaced in the wrong countries. 

Though our HLPNN model achieves great performances under Text+Net and Text+Meta+Net settings, potential improvements could be made using better graph-level classification frameworks. We currently only use network information to train network embeddings as user-level features. For future work, we would like to explore ways to combine graph-level classification methods and our user-level learning model. Propagating features from connected friends would provide much more information than just using network embedding vectors. Besides, our model assumes each post of one user all comes from one single home location but ignores the dynamic user movement pattern like traveling. We plan to incorporate temporal states to capture location changes in future work.

\section*{Acknowledgments}
This work was supported in part by the Office of Naval Research (ONR) N0001418SB001 and N000141812108. The views and conclusions contained in this document are those of the authors and should not be interpreted as representing the official policies, either expressed or implied, of the ONR.

\bibliography{acl2019}
\bibliographystyle{acl_natbib}

\end{document}